\documentclass{aa}
\usepackage{psfig,natbib,amssymb}

\bibpunct{(}{)}{;}{a}{}{,} 

\begin{document}
\headnote{Letter to the Editor} 

\title{The discovery of the ``21'' $\mu$m and ``30'' $\mu$m emission
  features in Planetary Nebulae with Wolf-Rayet central stars\thanks{
    based on observations obtained with ISO, an ESA project with
    instruments funded by ESA Member states (especially the PI
    countries: France, Germany, the Netherlands and the United
    Kingdom) with the participation of ISAS and NASA}}

\author{ S. Hony\inst{1}, L.B.F.M. Waters\inst{1,2}, A.G.G.M.
  Tielens\inst{3,4} }

\authorrunning{Hony et al.}
\titlerunning{``21'' and ``30'' $\mu$m features in [WC]-PNe}

\institute{
  Astronomical Institute ``Anton Pannekoek'', University of Amsterdam,
  Kruislaan 403, NL--1098 SJ Amsterdam, The Netherlands
  \and
  Instituut voor Sterrenkunde, Katholieke Universiteit Leuven,
  Celestijnenlaan 200B, B--3001 Heverlee, Belgium
  \and
  SRON Laboratory for Space Research Groningen, 
  P.O. Box 800, NL--9700 AV Groningen, The Netherlands 
  \and
  Kapteyn Astronomical Institute PO Box 800, NL--9700 AV
  Groningen, The Netherlands}

\offprints{Sacha Hony (hony@astro.uva.nl)}
\date{Received $<$date$>$; accepted $<$date$>$}

\abstract{We report the discovery of the ``21'' $\mu$m and ``30''
  $\mu$m features in the planetary nebulae around the
  hydrogen-deficient stars HD~826 and HD~158269. The carriers of these
  features are known to be produced in outflows around carbon-rich
  stars.  This discovery demonstrates that the bulk of the dust in
  these nebulae has been produced during a carbon-rich phase before
  the atmospheres of these stars became hydrogen poor. This is the
  first time that the ``21'' $\mu$m feature has been detected in any
  planetary nebula. It shows that once formed its carrier can survive
  the formation of the nebula and the exposure to the UV radiation of
  the hot central star. This means that the carrier of ``21'' $\mu$m
  feature is not transient: the absence of the feature sets limits on
  the production of its carrier.  
  \keywords{Circumstellar matter -- planetary nebulae: individual:
    NGC 40, NGC 6369 -- Stars: mass-loss -- Stars: evolution } }

\maketitle

\section{Introduction}
The precise evolutionary channel leading to the formation of Planetary
Nebulae (PNe) with Wolf-Rayet ([WC]) central stars ([WC]-PNe) is not
well established. While abundance studies of central stars of [WC]-PNe
indicate that they are very H-poor, the nebulae show ample evidence
for H-rich gas \citep[e.g.][]{1995A&A...303..893G} and dust.  Model
calculations of AGB evolution show that the transition from H-rich to
H-poor may be the result of a thermal pulse (TP) when the envelope
mass has dropped below some threshold value, which may be as high as
$\sim$0.01 M$_{\odot}$ \citep{1999A&A...349L...5H}. It is not well
understood whether this TP occurs when the star is still on the AGB or
during the post-AGB evolution. For some stars, like \object{V605 Aql},
this transition has clearly occurred during the post-AGB phase. These
systems harbour a small H-poor nebula within a large H-rich old nebula
\citep[e.g.][]{1992MNRAS.257P..33P}.  The stellar C~{\sc iv} emission
line in V605~Aql implies a [WC] spectral type
\citep{1996ApJ...472..711G}. On the other hand, statistical analysis
of the [WC]-PNe show that they are as a group not older than other PNe
\citep{2001Ap&SS.275...67G} and thus most of their central stars must
have lost their H-rich envelope during or very soon after the AGB.

We are studying the infrared spectra of [WC]-PNe in order to
reconstruct the mass loss history and chemical evolution of these
objects.  The dust content not only traces the conditions during the
preceding phases; links them to precursor objects but also shows the
type of materials fed into the ISM. The IR spectral appearance of
C-rich evolved stars differs strongly from AGB, post-AGB to PNe
objects. This is often interpreted as evidence for transient dust
components.

In this \emph{Letter} we report on the discovery of the ``21'' $\mu$m
and ``30'' $\mu$m dust features in two [WC]-PNe (\object{NGC 40}
([WC~8]) and \object{NGC 6369} ([WC~4])). These are ascribed to C-rich
dust. The detection of the ``21'' $\mu$m feature is the first in a PN.
We show that these detections imply that the TP which converted these
stars to C-rich objects did not remove all H from the atmosphere of
the star and was not responsible for the termination of the AGB.
Furthermore, these observations may establish an evolutionary link
between the [WC]-PNe and the ``21'' $\mu$m emitting post-AGB objects
with cool central stars.

\section{The observations}
\begin{table}
  \caption{
    Source list. Observational details of the sources used in this study.}
  \label{tab:obslog}
  \begin{tabular}{r@{\hspace{4pt}} r@{\hspace{4pt}} r@{\hspace{4pt}} l@{\hspace{4pt}} l@{\hspace{4pt}}}
    \hline
    \hline
\multicolumn{1}{c}{$\alpha$(J2000)}&
\multicolumn{1}{c}{$\delta$(J2000)}& 
\multicolumn{1}{c}{TDT$^{a}$}&
                   obs.mode$^{b}$&
                   obs.id\\
\hline
\multicolumn{5}{c}{\bf NGC 40; HD~826; IRAS~00102$+$7214} \\
 00\,13\,01.10 & $+$72\,31\,19.09 & 30003803& SWS01(3)& SWS\_CAL \\
 00\,13\,00.91 & $+$72\,31\,19.99 & 44401917& SWS01(2)& MBARLOW  \\
 00\,13\,01.10 & $+$72\,31\,19.09 & 81101203& SWS06   & SWS\_CAL \\
 00\,13\,00.91 & $+$72\,31\,19.99 & 47300616& LWS01   & MBARLOW  \\
\hline
\multicolumn{5}{c}{\bf NGC 6369; HD 158269; IRAS 17262$-$2343} \\
 17\,29\,20.80 & $-$23\,45\,32.00 & 45601901& SWS01(1)&  SGORNY  \\
 17\,29\,20.80 & $-$23\,45\,32.00 & 31100910& LWS01   &  CZHANG  \\
\hline
\hline
\end{tabular}
$^{a}$TDT number which uniquely identifies each ISO observation.\\
$^{b}$SWS/LWS observing mode used \citep[see][]{1996A&A...315L..49D,
  1996A&A...315L..38C}. Numbers in brackets correspond to the scanning
speed.\\
\end{table}

The data were obtained using the Short Wavelength Spectrometer (SWS)
\citep{1996A&A...315L..49D} on-board the Infrared Space Observatory
(ISO) \citep{1996A&A...315L..27K}. Details on the observations are
given in Table~\ref{tab:obslog}.

The data were processed using SWS interactive analysis product; IA$^3$
\citep[see][]{1996A&A...315L..49D} using calibration files and
procedures equivalent to pipeline version 9.5. NGC~40 has been
observed multiple times and the data have been co-added after the
pipeline reduction and bad data removal. Since the features we discuss
here are fully resolved in all observing modes, we can safely combine
the data obtained in all different modes maximising the S/N. Further
data processing consisted of rebinning on a fixed resolution
wavelength grid. The match between the individual sub-bands is
excellent for both sources and there is no need to splice the
sub-bands.

We compare the SWS spectra with the available IRAS photometry and the
IRAS/LRS spectra. We find that the IRAS photometry lies well above the
SWS observations. This indicates that the sources are more extended
than the SWS apertures. Indeed the optical nebulae associated with
NGC~40 and NGC~6369 are 60\arcsec\ and 30\arcsec\ while
the largest SWS aperture measures 20$\times$33\arcsec.  Surprisingly
the shape and slope of the SWS and IRAS/LRS spectra correspond very
well. This indicates that the SWS spectrum is representative of the
\emph{bulk} of the dust in those nebula even if we do not see
\emph{all} dust emission.

Extended sources give rise to aperture jumps between the SWS sub-bands
most notably around 29 $\mu$m because at those wavelengths the
effective aperture changes from 14$\times$27\arcsec\ to
20$\times$33\arcsec. We do not observe a flux jump in NGC~6369 and at
most a jump of 20 percent in NGC~40. This means that there is not more
dust located in the larger aperture. NGC~6369 has a ring-like
structure in the optical.  The SWS spectrum was taken from the east
part of this ring and the largest aperture sees more of the relatively
empty inner part of the ring.

The LWS spectra are observed through a circular aperture of 1\arcmin.
We do see a flux jump between the SWS and LWS spectra. The factors
needed to bring the SWS and LWS spectra together at 45 $\mu$m are the
same needed to bring the SWS and the IRAS/LRS spectra together, 
demonstrating that the LWS instrument does see the entire nebula.

\section{Description of the spectra}
\begin{figure}[t]
  \psfig{figure=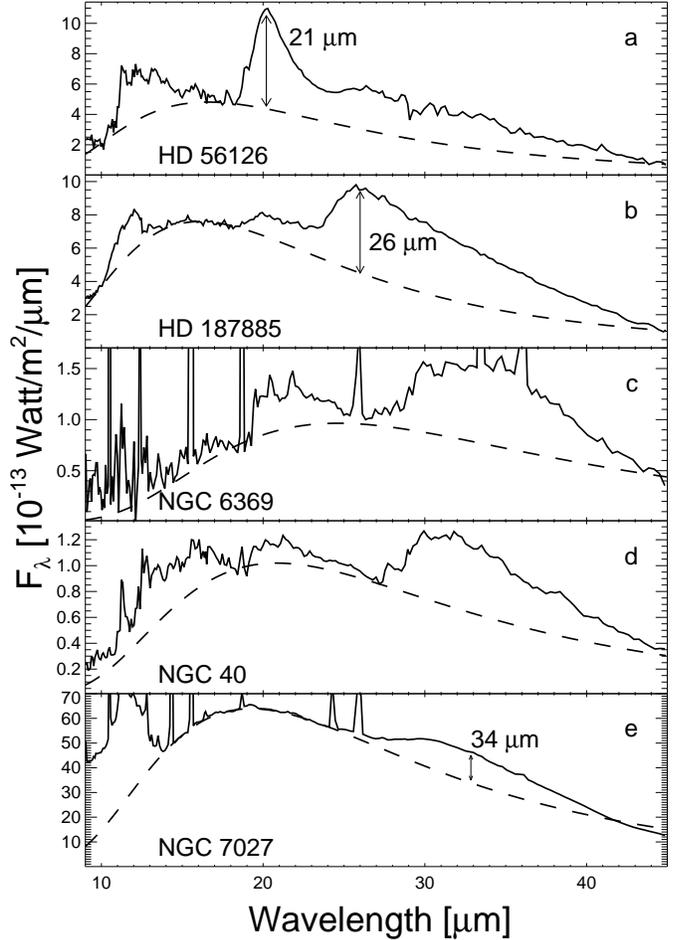,width=8.8cm}
  \caption{Overview of the spectra of NGC~40 and NGC~6369. For
    comparison we also show the spectra of HD~56126, HD~187885 and
    NGC~7027. The dashed lines represent the continua we draw.}
  \label{fig:fig1}
\end{figure}
We show the observed spectra of NGC~40 and NGC~6369 from 10 to 45
micron in Fig.~\ref{fig:fig1}c,d. Both mid-IR spectra are dominated by
broad emission excesses from 20$-$25 and 26$-$45 $\mu$m. We also
detect emission from polycyclic aromatic hydrocarbons (PAHs)
\citep{ATB89} between 3.3$-$12.7 $\mu$m. NGC~40 also has a strong
unidentified emission plateau from 10$-$19 $\mu$m. We concentrate on
the two broad features around 20 and 30 $\mu$m. For comparison we also
show the SWS spectra of the 21 $\mu$m objects \object{HD 187885} and
\object{HD 56126} and the C-rich PN \object{NGC 7027}
(Fig.~\ref{fig:fig1}a,b,e, respectively).

A broad feature peaking near 21 $\mu$m is found to be an important
component of the IR spectra of some C-rich proto-planetary nebulae
(PPNe) \citep[e.g.][]{1989ApJ...345L..51K}. This feature has up to now
only been detected in a very uniform group of post-AGB sources;
C-rich, within a narrow temperature range \citep{1999IAUS..191..297K},
metal poor and s-process enhanced \citep{2000A&A...354..135V}. These
sources are generally termed the ``21 $\mu$m objects''.

A very broad emission feature extending from 25$-$45 $\mu$m is
abundantly detected in IR spectra of a variety of C-rich evolved
objects ranging from intermediate mass loss AGB stars to PNe
\citep{1981ApJ...248..195F}. The feature is tentatively ascribed to
MgS \citep[e.g.][]{1985ApJ...290L..41N, 1985ApJ...290L..35G,
  1994ApJ...423L..71B}. \citet{2000ApJ...535..275H} show that the
``30'' $\mu$m feature in the 21 $\mu$m objects consists of a 26 and a
30 $\mu$m component. An extensive survey of the ``30'' $\mu$m feature
in a wide range of C-rich evolved objects shows that its peak position
can vary from 26 $\mu$m to 34 $\mu$m (Hony et al., in prep), see
Fig.~\ref{fig:fig1}b,e for an example of this shift.  The two [WC]-PNe
show a peak near 34 $\mu$m as do most regular C-rich PNe.  In contrast
in the 21 $\mu$m objects the ``30'' $\mu$m feature usually peaks at 26
$\mu$m as it does in most AGB stars.

\begin{figure}[t]
  \psfig{figure=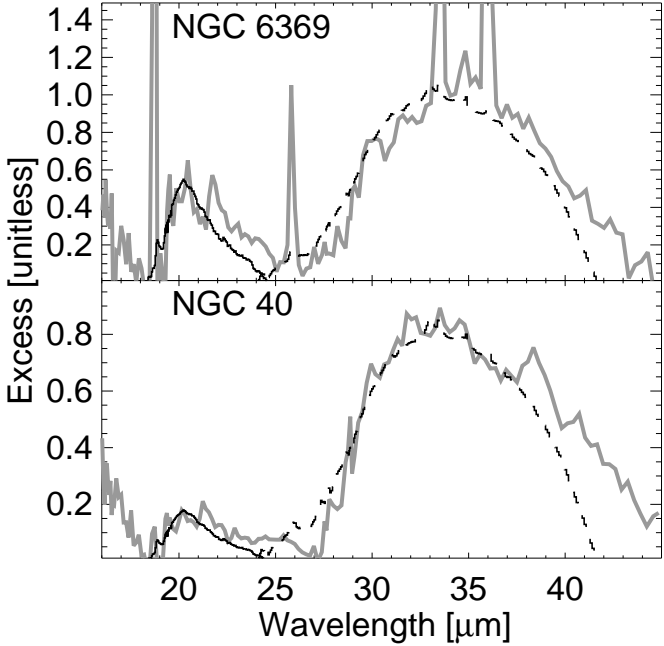,width=8.8cm}
  \caption{A comparison of the observed emission features compared
    with the ``21'' $\mu$m (solid line) and the ``30'' $\mu$m (dashed
    line) features as derived from HD~56126 and NGC~7027
    respectively.}
  \label{fig:fig2}
\end{figure}
To compare the features we observe in NGC~40 and NGC~6369 with other
sources we extract the profiles of the excess emission. We construct a
simple model continuum by fitting a modified blackbody
(F$_{\nu}$($\lambda$) =
B$_{\nu}$($\lambda$,T)$\times{\lambda}^{-p}$, where
B$_{\nu}$($\lambda$,T) is the Planck function.) to continuum points
around 9$-$10, 16 and 45 $\mu$m. We keep the power ($p$) fixed at 0.5,
a value which fits those sources well and also matches the slope of
the LWS spectra when available. These continua are shown as dashed
lines in Fig.~\ref{fig:fig1}.  The excess emission in both sources is
shown in Fig.~\ref{fig:fig2}.

For comparison we show the ``30'' $\mu$m feature of the C-rich PN
NGC~7027 in Fig.~\ref{fig:fig2}. The ``30'' $\mu$m feature in the two
[WC]-PNe compares well, both in peak position and in width, with the
other PNe.  The discrepancies at the longest wavelength may reflect
the difficulties to draw a continuum for the spectrum of NGC~7027. The
SWS and LWS spectra of NGC~7027 show a depression around 42 $\mu$m.
This might be due to a residual instrumental effect.

For the comparison of the ``21'' $\mu$m feature in the [WC]-PNe with
that in 21 $\mu$m objects, we have decomposed the emission from the
latter objects in a 21 $\mu$m and a 26 $\mu$m component (see also
Fig.~\ref{fig:fig1}).  The thus derived ``21'' $\mu$m component rises
steeply from 18.4 to 20.1 $\mu$m and slowly tapers off until 24.6
$\mu$m. See also \citep{1999ApJ...516L..99V} who show the profile to
be identical in all known 21 $\mu$m sources. We compare this profile
to the excesses found in NGC~40 and NGC~6369 in Fig.~\ref{fig:fig2}.
The profile compares well with the new observations, specifically the
sharp rise and the long tail towards longer wavelengths. We also find
an extra contribution near 21.5 $\mu$m not found in the 21 $\mu$m
objects.  This is the first time the ``21'' $\mu$m feature has been
detected in an object other than a PPN with a low mass progenitor.

\section{Discussion}
The ``21'' $\mu$m feature is absent during the AGB phase, therefore
its carrier is generally thought to have formed during a short period
of enhanced mass loss which terminated the AGB phase.  However this
scenario does not explain why it is not detected in their successors,
the PNe.  This has led to the suggestion that the carrier is a
transient, unstable constituent of the dust. These new detections
demonstrate that the ``21'' $\mu$m carrier can survive into the PN
phase, where it is exposed to UV radiation field of the hot central
star; \emph{it is a non-transient, stable dust component.}
\citet{2000Helden} identified TiC as the carrier of the ``21'' $\mu$m
feature. The non-transient nature of the feature further strengthens
its identification with the refractory TiC. This identification
however implies extreme conditions during formation with high
densities and consequently very high mass-loss rates: typically
$>$10$^{-4}$ M$_{\odot}$/yr.

The detection of these features in the spectra of [WC]-PNe has
ramifications for our understanding of their evolution. The ``30''
$\mu$m feature is characteristic for the C-rich ejecta of AGB and
post-AGB objects. The dominance of this band in the spectra of these
[WC]-PNe implies that the bulk of the dust in these nebula was formed
during a preceding C-rich phase.  Clearly, the transition from O-rich
to C-rich and the transition from H-rich to H-poor are decoupled for
these objects.  Thus, these nebulae have undergone normal AGB
evolution from O-rich to C-rich well before the loss of the last 0.01
M$_{\odot}$ turned the central star into a [WC] star.

This seems to be a very general characteristic for [WC]-PNe.  The
mid-IR spectra of all of these objects are dominated by the well known
emission features due to PAHs \citep{1989ApJ...344L..13C,
  1999ApJ...513L.135C}. The strong 3.3, 8.6, and 11.3 $\mu$m bands in
these spectra attests to the presence (and importance) of H during the
formation of the PAHs.  In contrast, population I WC stars show mid IR
spectra characterised by strong continua with very weak absorption
features at 6.2 and 7.7 without any sign of the 3.3 and 11.2 $\mu$m
feature \citep{2001ApJ...550L.207C}. We conclude that the [WC]-PNe
went through a H-rich, C-rich phase during which the PAHs condensed.

The ```21'' $\mu$m feature has been detected in C-rich post-AGB
objects, however not in their precursors, the carbon stars.  Hence,
the the ``21'' $\mu$m carrier have must formed at the end of the AGB,
during a burst of mass loss, which lasted short compared to the
post-AGB phase, i.e., $\lesssim$ 1000 years. If we assume that the
``21'' $\mu$m carrier in the [WC]-PNe nebulae condensed similarly to
all other known 21 $\mu$m emitting sources, this means this mass loss
burst occurred \emph{prior} to the TP which turned the star H-poor.
Between leaving the AGB and this TP these objects may have appeared
similar to the 21 $\mu$m objects.  Likewise if a cool 21 $\mu$m object
were to suffer a TP -- turning it H-poor -- it would become a PN
similar to these PNe. We conclude that NGC~40 and NGC~6369 could
represent successors to some of the cool 21 $\mu$m objects and that the
\emph{last} TP followed the mass loss burst. The low dynamical ages of
the nebulae of NGC~40 and NGC~6369 ($\sim$5000 and $\sim$1500 yr) and
the much longer timescale for TPs ($\gtrsim$10$^4$ yr) shows that the
production of the ``21'' $\mu$m carrier is not triggered by a TP.

It is tempting to further explore the evolutionary link between the
``21'' $\mu$m emitting PPNe and the [WC]-PNe, given the fact that
these PPNe are the only objects known to exhibit the feature apart
from the two PNe we present here. \citet{2000A&A...354..135V} show
that the 21 $\mu$m objects as a group are metal poor albeit with large
intrinsic range ([Fe/H]=$-$0.3 to $-$1.0).  Abundance determinations of
NGC~40 and NGC~6369 \citep{1991ApJS...76..687P} show that [Ne/H] is
$-$0.8 and $-$0.5. Assuming that the Ne abundance in those PNe is
representative for the total metallicity \citep[e.g.][chapter
III.E]{1984assl..107.....P}, this means they fall in the range also
observed for the 21 $\mu$m objects.

It is important to note that there cannot be a one-to-one
correspondence between the 21 $\mu$m objects and the [WC]-PNe. Many
[WC]-PNe show \emph{no} signature of a long lasting C-rich mass-loss
phase. Rather these nebulae are typified by a mixed chemistry with
warm ejecta from both an O-rich and a C-rich phase
\citep{1998A&A...331L..61W}. This is interpreted as evidence for a
short ($\sim$1000 yr) transition phase during which the star first
became C-rich and shortly after H-poor
\citep{1998A&A...331L..61W,1999ApJ...513L.135C}, because the O-rich
ejecta predate the C-rich material while still being close to the
star. The ``21'' $\mu$m feature is not present in the mid-IR spectra
of these mixed chemistry sources, excluding them as successors to the
21 $\mu$m objects.  Likewise not every 21 $\mu$m object may evolve to
become H-poor.  Whether a star becomes H-poor is determined by the
remaining envelope mass at the time of the last TP. Thus, the question
arises where the PNe with the H-rich central stars and the ``21''
$\mu$m feature are. Due to observer bias few such PNe have been
observed with SWS, however those which have been studied show no
``21'' $\mu$m feature. We now know that the feature is not transient.
\emph{Thus the absence of the feature implies non-production of its
  carrier}.  Hence, those objects have not gone through a phase in
which they produced the ``21'' $\mu$m carrier. We conclude that the
chemical evolution, as determined by the H content of the atmosphere
of the central star, and the mass-loss history; specifically the mass
loss burst as traced by the ``21'' $\mu$m feature, are decoupled.

\acknowledgements{LBFMW and SH acknowledge financial support from an
  NWO \emph{Pionier} grant.}

\bibliographystyle{apj}
\bibliography{articles}
\end{document}